\documentclass[preprint,showpacs,preprintnumbers,amsmath,amssymb,apl]{revtex4-1}

\usepackage[dvipdfmx]{graphicx}
\usepackage{dcolumn}
\usepackage{bm}
\usepackage{color} 
\begin{document}
%
\title{Equivalent circuit representation of hysteresis in solar cells that considers 
interface charge accumulation: Potential cause of hysteresis in perovskite solar cells}
\author{
Kazuhiko Seki
}
\email{k-seki@aist.go.jp}
\affiliation{Nanomaterials Research Institute(NMRI), 
National Institute of Advanced Industrial Science and Technology (AIST)\\
AIST Tsukuba Central 5, Higashi 1-1-1, Tsukuba, Ibaraki 305-8565, Japan 
}

\date{\today}
\begin{abstract}
If charge carriers accumulate in the charge transport layer of a solar cell, 
then the transient response of the electric field that originates from these accumulated charges 
results in hysteresis in the current-voltage ($J$-$V$) characteristics.  
While this mechanism was previously known, 
a theoretical model to explain these $J$-$V$ characteristics 
has not been considered to date. 
We derived an equivalent circuit from the proposed hysteresis mechanism. 
By solving the equivalent circuit model, we were able to 
reproduce some of the features of hysteresis in perovskite solar cells. 
\end{abstract}

\maketitle
Recently, organic-inorganic hybrid-perovskite solar cells have attracted considerable research attention because of the rapid increase in 
the power conversion efficiency that can be gained by avoiding the use of electrolytes. \cite{Green_14}
Perovskite solar cells can be fabricated using solution-based processes and low-cost materials, which gives them an advantage in the solar cell market. \cite{Green_14} 
The power conversion efficiency of these devices is more than $20$ \% 
\cite{meloni16} and is approaching the theoretically estimated limit. \cite{Seki15}
At present, perovskite solar cells are considered to be neither $p-n$ junction-type solar cells nor excitonic-type solar cells like organic solar cells. 
The standard structure of a perovskite solar cell includes a charge transport layer, 
where holes or electrons are selectively transported to the appropriate electrode, thus suppressing recombination losses. 
Perovskite solar cells have unique features that are not found in conventional solar cells. 
However, unresolved issues remain with regard to their unique properties. 
One major issue for perovskite solar cells is the hysteresis that appears in their
electrical characteristics. \cite{Snaith14,meloni16,Tress_15,Unger14,Zhang_15,Chen16,Jena16,Cojocaru15,Ono15,Nagaoka15,Tao15}

To aid in understanding of the electrical processes in a solar cell, 
the current density, $J$, is studied by varying the applied voltage, $V$. 
The maximum power conversion efficiencies (PCEs) of the devices 
can then be obtained from these $J$-$V$ characteristics. 
In perovskite solar cells, 
hysteresis is often observed in the $J$-$V$ characteristics.
If hysteresis is present, 
then it becomes difficult to identify the maximum power point. \cite{Snaith14,Zhang_15,Jena16}

In attempts to determine the possible origin of the hysteresis behavior, 
several potential mechanisms have been proposed. \cite{Snaith14,Zhang_15,Jena16,Chen16,Cojocaru15,Ono15,meloni16,Zhou_15,Nagaoka15,Tao15,Unger14}
Some experiments have shown that 
the hysteresis is correlated with the ferroelectric characteristics of the perovskite materials.  
\cite{Snaith14,Zhang_15,Unger14,Zhou_15}
The internal electric field that is associated with the 
movement of cations or anions in the perovskite material is also considered to be related to the hysteresis.  \cite{Snaith14,Zhang_15,meloni16,Chen16,Zhou_15}
In this work, we consider a mechanism by 
the electric field from the accumulated charges in the charge transport layer. \cite{Chen16,Jena16,Cojocaru15,Ono15,Nagaoka15,Tao15}
While this mechanism was previously known, 
a theoretical model to explain the $J$-$V$ characteristics of the device 
has not previously been considered. 
This mechanism could also be applied in other situations where charge accumulation occurs within the charge generation layer. In this case, the accumulated charges at the interface can be ions.

In equivalent circuit models of solar cells, the dark current is generally represented using a diode equation that was given by \cite{sze06}:
$J_d= J_0 \left[
\exp\left[qV_d/(n k_{\rm B} T)
\right]-1
\right], 
$ where $q$, $J_0$, and $V_d$ are the elementary charge, the reverse saturation current density and the applied voltage across 
the diode, respectively; $k_B$ is the Boltzmann constant, $T$ is the temperature, and $n$ is a phenomenological constant that is equal to one for an ideal diode. 
When there is a photo-generated current density, $J_{\rm ph}$, 
the total current density in the circuit element is then given by \cite{sze06}
\begin{align}
J=J_{\rm ph}-J_0 \left[
\exp\left(\frac{q(V+R_{\rm s}J)}{n k_{\rm B} T}
\right)-1
\right] -\frac{V+R_{\rm s}J}{R_{\rm sh}},
\label{eq:solarcellsEC}
\end{align}
where $V$ is the external voltage across the output terminal, and $R_{\rm s}$ and $R_{\rm sh}$ represent the series and shunt resistances, respectively. 
The corresponding 
circuit is shown in Fig. 1 (a).  
\begin{figure}
\centerline{
\includegraphics[width=0.6\columnwidth]{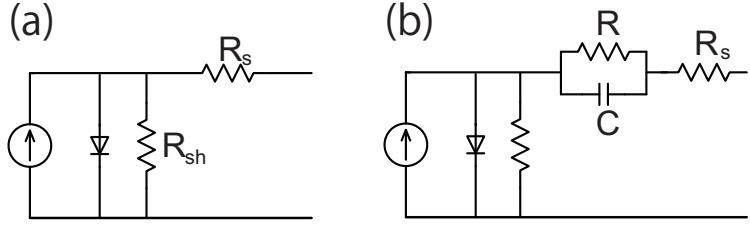}
}
\caption{Equivalent circuit representing voltage reduction caused by charge accumulation at the interfaces: 
(a) conventional electric circuit model for solar cells; (b) proposed model used in this study.}
\label{fig:1}
\end{figure}

We consider the case where an additional resistor-capacitor ($R$-$C$) unit is inserted into the circuit, as shown in Fig. \ref{fig:1}(b). 
The additional $R$-$C$ circuit represents the voltage reduction that occurs through charge accumulation at the interfaces. 
The source of this voltage reduction may be the screening of the electric fields by the accumulated charges; 
this will be explained later in the paper. 
When the voltage applied to the $R$-$C$ circuit, the external voltage across the output terminals, 
and the voltage across the element are denoted by $V_i(t)$, $V(t)$,
and $V_d (t)$, respectively, these parameters are related by the following equation: 
\begin{align}
V_d (t)=V(t)+V_i(t)+R_{\rm s} J,  
\label{eq:Vbalance}
\end{align}
where the current density is given by 
\begin{align}
J=J_{\rm ph}-J_0 \left[
\exp\left(\frac{q V_d(t)}{n k_{\rm B} T}
\right)-1
\right] -\frac{V_d(t)}{R_{\rm sh}}.  
\label{eq:Jin}
\end{align}
The current conservation law leads to 
\begin{align}
C\frac{d}{dt} V_i(t)+\frac{1}{R} V_i(t) =J (t).
\label{eq:Cconservation}
\end{align}
We then substitute Eqs. (\ref{eq:Vbalance}) and (\ref{eq:Jin}) 
into Eq. (\ref{eq:Cconservation}) to simulate the $J$-$V$ curves by setting $n=1$. 
In the steady state, the solution can be reduced to Eq. (\ref{eq:solarcellsEC}), with 
$R_{\rm s}+R$ in the place of $R_{\rm s}$. 


\begin{figure}
\centerline{
\includegraphics[width=0.6\columnwidth]{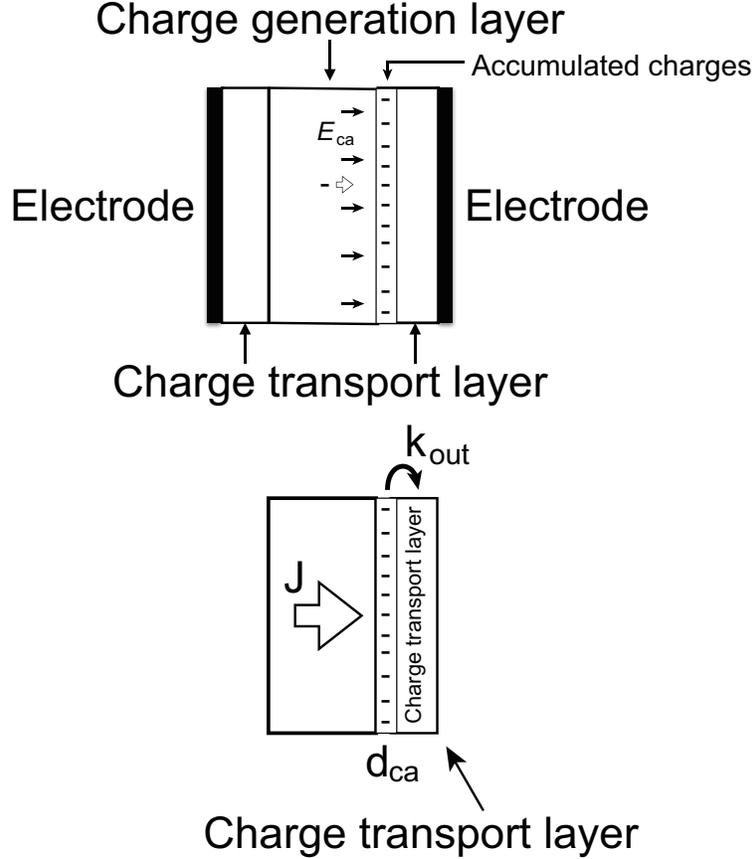}
}
\caption{Schematic illustration of charge accumulation at the interface between the charge generation layer and 
the charge transport layer on the right-hand side. 
The accumulated charges create an electric field in the charge generation layer 
that opposes the flow of charge. (An electron experiences a force acting in the direction opposite to that of the field vector.)
}
\label{fig:2}
\end{figure}

To provide a simple physical description, 
we consider the case where 
holes and electrons are generated in the charge generation layer and the carriers are 
transferred to each charge transport layer 
as shown in Fig. \ref{fig:2}. \cite{Jena16,Zhou_15} 
If these charge carriers accumulate at the interface between the charge generation layer and 
the charge transport layer on the right-hand side of the structure because 
the charge transfer rate to the charge transport layer is low, \cite{Jena16,Zhou_15}
the carrier density in the charge accumulation region, 
which is represented by $\rho_{\rm ca}$, is given by the following rate equation:
\begin{align}
\frac{d}{dt} \rho_{\rm ca} d_{\rm ca}= \theta J - k_{\rm out} \rho_{\rm ca} d_{\rm ca},
\label{eq:rate}
\end{align}
where $k_{\rm out}$ and $d_{\rm ca}$ represent the rate constant for charge carrier transfer 
from the charge accumulation region to the charge transport layer and 
the depth of the charge accumulation region, respectively. 
$J$ represents the current that flows out from the charge generation layer, as given by 
Eq. (\ref{eq:Jin}). 
$\theta$ is equal to $1-$ the transmission coefficient of the current flow.
If all the currents accumulate, we have $\theta=1$. 
We consider the case where $d_{\rm ca}$ is smaller than the characteristic length of the concentration gradient,  
which means that the surface charge density 
$\sigma_{\rm ca}$ can be given by $\sigma_{\rm ca}=\rho_{\rm ca}d_{\rm ca}$. 
When the dielectric constants of the charge generation layer and the charge transport layer are denoted 
by $\epsilon_{\rm cg}$ and $\epsilon_{\rm ct}$, respectively, 
the strength of the electric field from the accumulated charges is then given by
$(\epsilon_{\rm cg}+\epsilon_{\rm ct}) E_{\rm ca}=\sigma_{\rm ca}$ using Gauss's theorem. 
In Eq. (\ref{eq:Jin}), 
$V_d$ represents the voltage that is applied to the charge generation layer. 
The accumulated charges at the interface produce an electric field in the charge generation layer 
in a direction such that it opposes the carrier flow. \cite{Zhou_15}
When a uniform electric field $E_{\rm ca}$ is applied to a charge generation layer 
of thickness $L$ in addition to the external voltage $V(t)$, 
then the net voltage that is applied to the charge generation layer is 
$V_d=V(t)+V_i$, 
where $V_i=E_{\rm ca} L$. 
By defining $R=L\theta/[k_{\rm out} (\epsilon_{\rm cg}+\epsilon_{\rm ct})]$ and $C=(\epsilon_{\rm cg}+\epsilon_{\rm ct})/(L\theta)$, 
Eq. (\ref{eq:rate}) can be reduced to Eq. (\ref{eq:Cconservation}), and 
the time constant that represents the transient response is given by 
$RC=1/k_{\rm out}$. 
Therefore, we can conclude that 
the equivalent circuit model shown in Fig. \ref{fig:1}, while approximate, 
represents the reduction of the electric field caused by charge accumulation 
at the interface between the charge generation layer and the 
charge transport layer. 
\begin{figure}
\centerline{
\includegraphics[width=0.4\columnwidth]{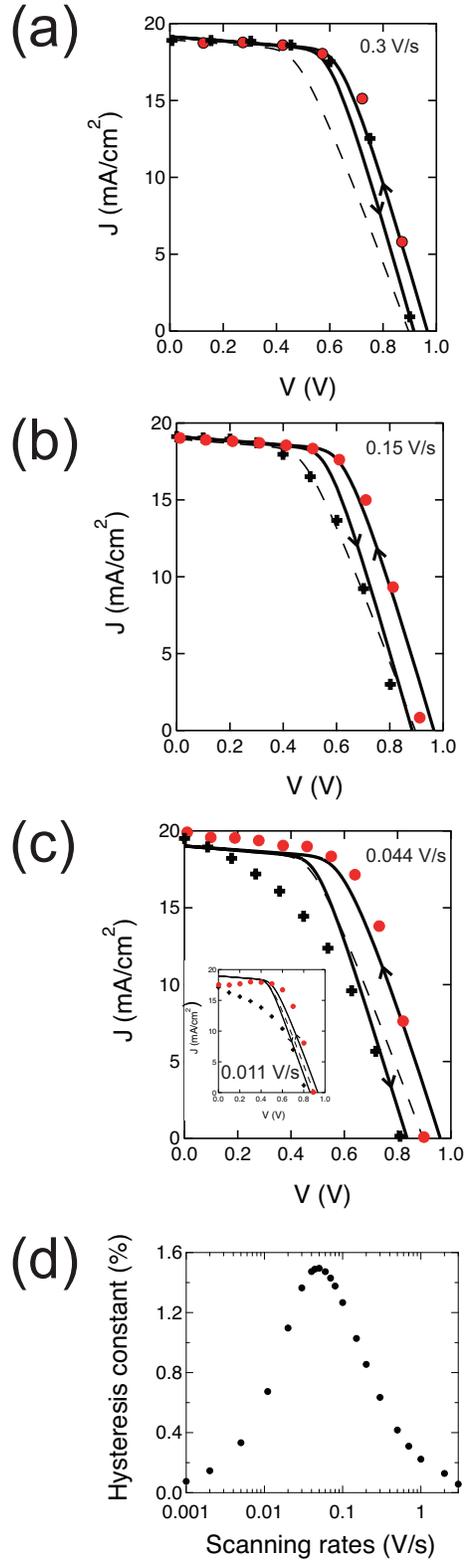}
}
\caption{(Color online) 
Current density as a function of external applied voltage. 
The experimental results of ref. \onlinecite{Snaith14} are shown 
as red circles (scanned from open-circuit) and black crosses (scanned from short-circuit), respectively. 
The scan rates are   
(a) 0.3 V/s, (b) 0.15 V/s, and (c) 0.044 V/s, while that in the inset of (c) is 0.011 V/s. 
Arrows indicate the scan directions. 
The dashed lines indicate the steady state result. 
(d) Hysteresis constant as a function of scan rate. 
}
\label{fig:3}
\end{figure}
\begin{figure}
\centerline{
\includegraphics[width=0.6\columnwidth]{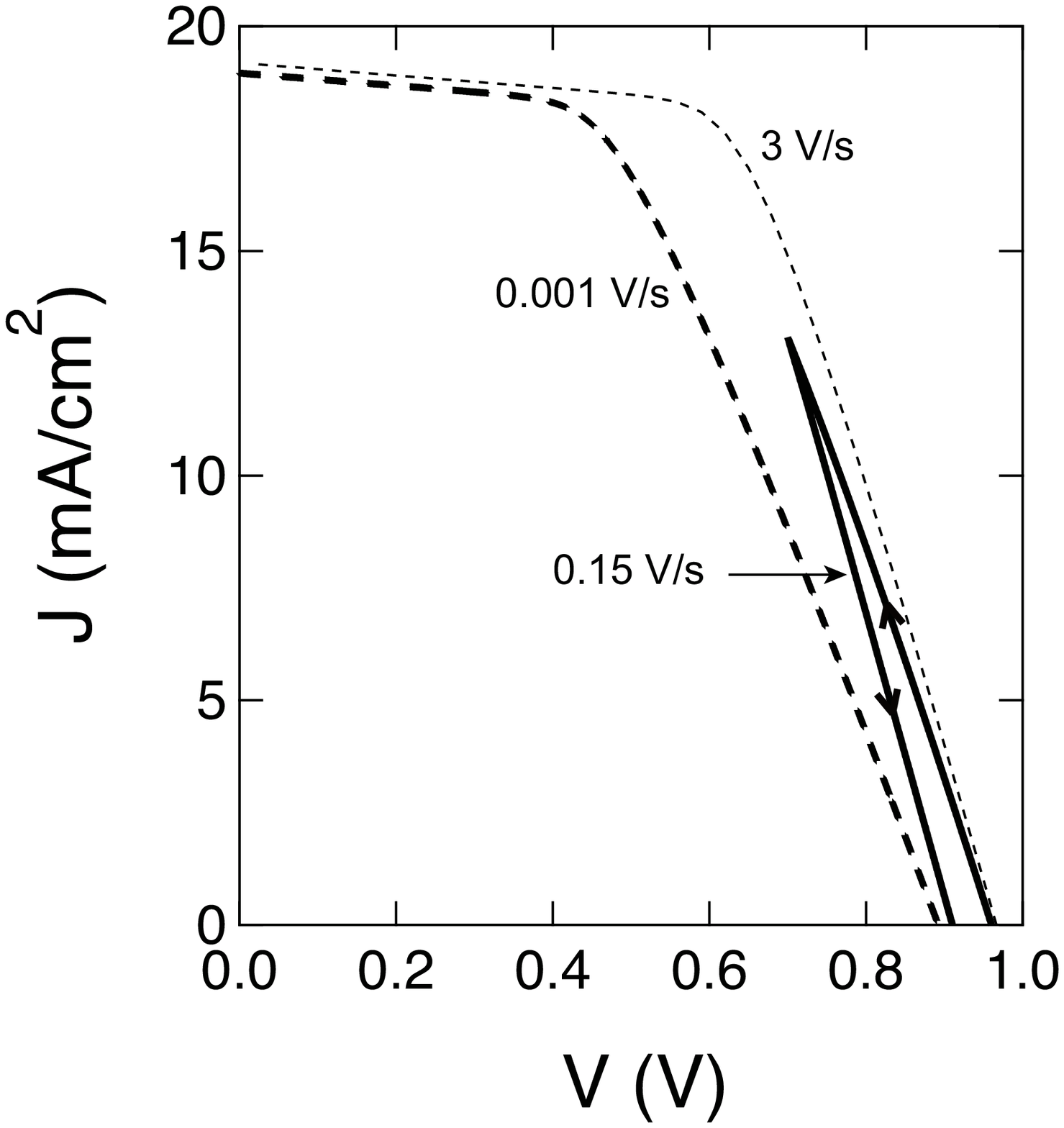}
}
\caption{
Current density as a function of external applied voltage 
when the amplitude of the voltage variation is small. 
(The scan rate is $0.15$ V/s.)
Arrows indicate the scan directions. 
$J$-$V$ characteristics are also shown for conditions where the scan rate is extremely fast and where it is extremely slow. 
}
\label{fig:4}
\end{figure}

Figure \ref{fig:3} shows the current density as a function of the external applied voltage at various 
scan rates. 
We consider the situation where
the applied voltage is maintained at 1.4 V for the first $10$ s, is then reduced to zero, and is reversed until 
the voltage returns to 1.4 V. 
The theoretical results are compared with the experimental results for planar heterojunction solar cells  
that are composed of layers of 
dense TiO$_{2}$ (electron transport layer), perovskite and 
Spiro-OMeTAD (hole transport layer).\cite{Snaith14}
In the numerical calculations, we set $T=300$ K, $J_{\rm ph}=19.5$ mA/cm$^2$, 
$J_0=0.5 \times 10^{-14}$ mA/cm$^2$, $r_{\rm sh}=700$ $\Omega$ cm$^2$, 
$R=5$ $\Omega$ cm$^2$, $C=3$ F/ cm$^2$, and $R_{\rm s}=15$ $\Omega$ cm$^2$. 

In the experiments, the hysteresis is higher when the voltage is scanned from open-circuit (OC) conditions than 
when initially scanned from the short-circuit (SC) conditions. \cite{Snaith14,Zhang_15,Jena16,Chen16,Cojocaru15,Ono15,meloni16,Nagaoka15,Unger14,Liu15,Kim_14}
This trend was reproduced theoretically. 
The theoretical results also show that
the hysteresis is greater on the OC side when compared 
with the SC side, as is commonly observed in planar cells. \cite{Tao15,Nagaoka15,Snaith14,Zhang_15,Jena16,Chen16,Liu15}
(In supplemental figure S1, we show an improved comparison to experiment.)\cite{SI}
Recently, to reduce or even to eliminate this hysteresis, the TiO$_2$ electron transport layer was replaced with another layer. 
\cite{*Tao15,Nagaoka15,Shao14,*Tripathi15}
The results presented here are consistent with the reduced hysteresis that was obtained by introducing these structures, 
because the charges would accumulate at the interface between the charge generation layer and the electron transport layer. 

The hysteresis was increased by reducing the scan rate from 0.3 to 0.044 V/s. 
When the scan rate was reduced further, 
the theoretical result shows a reduction in the hysteresis,  
but the hysteresis does not decrease in the experiment. 
The results are dependent on the experiments. 
In some experiments, the hysteresis was reduced, and the $J$-$V$ curve then approaches the steady state result 
when the scan rate is low. \cite{Zhou_15,Tress_15}
In Fig. \ref{fig:3}(d), we show the hysteresis constant as a function of scan rate. 
The hysteresis constant was defined 
as the difference between the maximum PCEs for the different scan directions
when assuming irradiation of $100$ mW cm$^{-2}$.  
The hysteresis constants have a maximum as a function of the scan rate,
and a similar maximum was recently reported. \cite{Zhou_15}
The scan rate at the maximum is close to the inverse of the relaxation time given by $RC=15$ s. 
If multiple relaxation times exist, the peak can then be broader and 
any reduction of the hysteresis produced by reducing the scan rate could hardly be observed. 

The hysteresis was reduced by scanning the voltage at either an extremely slow speed or a high speed, 
as observed experimentally. 
\cite{Snaith14,Zhang_15,Unger14,Zhou_15,Chen16,Tress_15}
In Fig. \ref{fig:4}, we show the results when the scan rate is extremely fast and when it is extremely slow. 
When the scan rate is low, the $J$-$V$ curve approaches the steady state result. \cite{Unger14,Zhou_15,Tress_15}
At the opposite limit of the fast scan rate, the fill factor is high. 
The accumulated charge density is low at the OC side when compared with that at the SC side. 
When the voltage scan starts at the OC side and the scan rate is too fast 
for charge accumulation to occur, 
the current density reduction produced by the accumulated charges is suppressed. 

In Fig. \ref{fig:4}, 
we also show the results obtained when the voltage variation is small. 
Although the amplitude of the voltage variation is small, hysteresis was still obtained, as 
observed experimentally. \cite{Kim_14}

In our theory, $\theta$ is introduced to provide a simple physical description. 
The value of $\theta$ may be small.  
If we use $C=(\epsilon_{\rm cg}+\epsilon_{\rm ct})/(L\theta)$, and assume 
that the relative dielectric constant is $20-50$ and $L=100$ nm, 
$C$ is then estimated to be sub-$\mu$F/cm$^2$ if $\theta=1$. 
The reported values of $C$ were widely distributed, ranging 
from sub-$\mu$F/cm$^2$ to $50$ mF/cm$^2$. \cite{Kim_14,Zarazua16,*Bo15}
These values and the high value of $C$ that was used to draw Fig. \ref{fig:3} suggest 
that only a proportion of the charge carriers are accumulated and that $\theta <1$.

The hysteresis that occurs because of charge accumulation can be distinguished from that 
which occurs due to ferroelectric polarization reversal 
when the dielectric relaxation occurs sufficiently rapidly.
If the ferroelectric polarization reversal occurs quickly, 
the hysteresis will then disappear using a small amplitude of the voltage scan.
However, if the dielectric relaxation is slow, 
the induced surface charge in the dielectric causes 
hysteresis similar to that which occurs because of charge accumulation. 

While our results are qualitative,
they do clearly reproduce the basic hysteresis features, such as a higher current when scanned from the OC side
as compared with that obtained when scanned from the SC side, and the large hysteresis at the OC side when compared with that on the SC side 
for the planar structured cells. 
\cite{Tao15,Nagaoka15,Snaith14,Zhang_15,Jena16,Chen16}

\acknowledgments
The author would like to thank Dr. Tetsuhiko Miyadera, Dr. Said Kazaoui, and Dr. Koji Matsubara at AIST for many fruitful discussions.


\begin{thebibliography}{22}%
\makeatletter
\providecommand \@ifxundefined [1]{%
 \@ifx{#1\undefined}
}%
\providecommand \@ifnum [1]{%
 \ifnum #1\expandafter \@firstoftwo
 \else \expandafter \@secondoftwo
 \fi
}%
\providecommand \@ifx [1]{%
 \ifx #1\expandafter \@firstoftwo
 \else \expandafter \@secondoftwo
 \fi
}%
\providecommand \natexlab [1]{#1}%
\providecommand \enquote  [1]{``#1''}%
\providecommand \bibnamefont  [1]{#1}%
\providecommand \bibfnamefont [1]{#1}%
\providecommand \citenamefont [1]{#1}%
\providecommand \href@noop [0]{\@secondoftwo}%
\providecommand \href [0]{\begingroup \@sanitize@url \@href}%
\providecommand \@href[1]{\@@startlink{#1}\@@href}%
\providecommand \@@href[1]{\endgroup#1\@@endlink}%
\providecommand \@sanitize@url [0]{\catcode `\\12\catcode `\$12\catcode
  `\&12\catcode `\#12\catcode `\^12\catcode `\_12\catcode `\%12\relax}%
\providecommand \@@startlink[1]{}%
\providecommand \@@endlink[0]{}%
\providecommand \url  [0]{\begingroup\@sanitize@url \@url }%
\providecommand \@url [1]{\endgroup\@href {#1}{\urlprefix }}%
\providecommand \urlprefix  [0]{URL }%
\providecommand \Eprint [0]{\href }%
\providecommand \doibase [0]{http://dx.doi.org/}%
\providecommand \selectlanguage [0]{\@gobble}%
\providecommand \bibinfo  [0]{\@secondoftwo}%
\providecommand \bibfield  [0]{\@secondoftwo}%
\providecommand \translation [1]{[#1]}%
\providecommand \BibitemOpen [0]{}%
\providecommand \bibitemStop [0]{}%
\providecommand \bibitemNoStop [0]{.\EOS\space}%
\providecommand \EOS [0]{\spacefactor3000\relax}%
\providecommand \BibitemShut  [1]{\csname bibitem#1\endcsname}%
\let\auto@bib@innerbib\@empty
\bibitem [{\citenamefont {Green}\ and\ \citenamefont
  {Snaith}(2014)}]{Green_14}%
  \BibitemOpen
  \bibfield  {author} {\bibinfo {author} {\bibfnamefont {M.~A.}\ \bibnamefont
  {Green}}\ and\ \bibinfo {author} {\bibfnamefont {H.~J.}\ \bibnamefont
  {Snaith}},\ }\href {\doibase 10.1038/nphoton.2014.134} {\bibfield  {journal}
  {\bibinfo  {journal} {Nat. Photon.}\ }\textbf {\bibinfo {volume} {8}},\
  \bibinfo {pages} {506} (\bibinfo {year} {2014})},\ \bibinfo {note} {and
  references cited therein.}\BibitemShut {Stop}%
\bibitem [{\citenamefont {Meloni}\ \emph {et~al.}(2016)\citenamefont {Meloni},
  \citenamefont {Moehl}, \citenamefont {Tress}, \citenamefont
  {Franckevi\v{c}ius}, \citenamefont {Saliba}, \citenamefont {Lee},
  \citenamefont {Gao}, \citenamefont {Nazeeruddin}, \citenamefont
  {Zakeeruddin}, \citenamefont {Rothlisberger},\ and\ \citenamefont
  {G\"{a}tzel}}]{meloni16}%
  \BibitemOpen
  \bibfield  {author} {\bibinfo {author} {\bibfnamefont {S.}~\bibnamefont
  {Meloni}}, \bibinfo {author} {\bibfnamefont {T.}~\bibnamefont {Moehl}},
  \bibinfo {author} {\bibfnamefont {W.}~\bibnamefont {Tress}}, \bibinfo
  {author} {\bibfnamefont {M.}~\bibnamefont {Franckevi\v{c}ius}}, \bibinfo
  {author} {\bibfnamefont {M.}~\bibnamefont {Saliba}}, \bibinfo {author}
  {\bibfnamefont {Y.~H.}\ \bibnamefont {Lee}}, \bibinfo {author} {\bibfnamefont
  {P.}~\bibnamefont {Gao}}, \bibinfo {author} {\bibfnamefont {M.~K.}\
  \bibnamefont {Nazeeruddin}}, \bibinfo {author} {\bibfnamefont {S.~M.}\
  \bibnamefont {Zakeeruddin}}, \bibinfo {author} {\bibfnamefont
  {U.}~\bibnamefont {Rothlisberger}}, \ and\ \bibinfo {author} {\bibfnamefont
  {M.}~\bibnamefont {G\"{a}tzel}},\ }\href {\doibase 10.1038/ncomms10334}
  {\bibfield  {journal} {\bibinfo  {journal} {Nat. Commun.}\ }\textbf {\bibinfo
  {volume} {7}},\ \bibinfo {pages} {10334} (\bibinfo {year}
  {2016})}\BibitemShut {NoStop}%
\bibitem [{\citenamefont {Seki}\ \emph {et~al.}(2015)\citenamefont {Seki},
  \citenamefont {Furube},\ and\ \citenamefont {Yoshida}}]{Seki15}%
  \BibitemOpen
  \bibfield  {author} {\bibinfo {author} {\bibfnamefont {K.}~\bibnamefont
  {Seki}}, \bibinfo {author} {\bibfnamefont {A.}~\bibnamefont {Furube}}, \ and\
  \bibinfo {author} {\bibfnamefont {Y.}~\bibnamefont {Yoshida}},\ }\href
  {http://stacks.iop.org/1347-4065/54/i=8S1/a=08KF04} {\bibfield  {journal}
  {\bibinfo  {journal} {Jpn. J. Appl. Phys.}\ }\textbf {\bibinfo {volume}
  {54}},\ \bibinfo {pages} {08KF04} (\bibinfo {year} {2015})}\BibitemShut
  {NoStop}%
\bibitem [{\citenamefont {Snaith}\ \emph {et~al.}(2014)\citenamefont {Snaith},
  \citenamefont {Abate}, \citenamefont {Ball}, \citenamefont {Eperon},
  \citenamefont {Leijtens}, \citenamefont {Noel}, \citenamefont {Stranks},
  \citenamefont {Wang}, \citenamefont {Wojciechowski},\ and\ \citenamefont
  {Zhang}}]{Snaith14}%
  \BibitemOpen
  \bibfield  {author} {\bibinfo {author} {\bibfnamefont {H.~J.}\ \bibnamefont
  {Snaith}}, \bibinfo {author} {\bibfnamefont {A.}~\bibnamefont {Abate}},
  \bibinfo {author} {\bibfnamefont {J.~M.}\ \bibnamefont {Ball}}, \bibinfo
  {author} {\bibfnamefont {G.~E.}\ \bibnamefont {Eperon}}, \bibinfo {author}
  {\bibfnamefont {T.}~\bibnamefont {Leijtens}}, \bibinfo {author}
  {\bibfnamefont {N.~K.}\ \bibnamefont {Noel}}, \bibinfo {author}
  {\bibfnamefont {S.~D.}\ \bibnamefont {Stranks}}, \bibinfo {author}
  {\bibfnamefont {J.~T.-W.}\ \bibnamefont {Wang}}, \bibinfo {author}
  {\bibfnamefont {K.}~\bibnamefont {Wojciechowski}}, \ and\ \bibinfo {author}
  {\bibfnamefont {W.}~\bibnamefont {Zhang}},\ }\href {\doibase
  10.1021/jz500113x} {\bibfield  {journal} {\bibinfo  {journal} {J. Phys. Chem.
  Lett.}\ }\textbf {\bibinfo {volume} {5}},\ \bibinfo {pages} {1511} (\bibinfo
  {year} {2014})}\BibitemShut {NoStop}%
\bibitem [{\citenamefont {Tress}\ \emph {et~al.}(2015)\citenamefont {Tress},
  \citenamefont {Marinova}, \citenamefont {Moehl}, \citenamefont {Zakeeruddin},
  \citenamefont {Nazeeruddin},\ and\ \citenamefont {Gr{\"a}tzel}}]{Tress_15}%
  \BibitemOpen
  \bibfield  {author} {\bibinfo {author} {\bibfnamefont {W.}~\bibnamefont
  {Tress}}, \bibinfo {author} {\bibfnamefont {N.}~\bibnamefont {Marinova}},
  \bibinfo {author} {\bibfnamefont {T.}~\bibnamefont {Moehl}}, \bibinfo
  {author} {\bibfnamefont {S.~M.}\ \bibnamefont {Zakeeruddin}}, \bibinfo
  {author} {\bibfnamefont {M.~K.}\ \bibnamefont {Nazeeruddin}}, \ and\ \bibinfo
  {author} {\bibfnamefont {M.}~\bibnamefont {Gr{\"a}tzel}},\ }\href {\doibase
  10.1039/C4EE03664F} {\bibfield  {journal} {\bibinfo  {journal} {Energy
  Environ. Sci.}\ }\textbf {\bibinfo {volume} {8}},\ \bibinfo {pages} {995}
  (\bibinfo {year} {2015})}\BibitemShut {NoStop}%
\bibitem [{\citenamefont {Unger}\ \emph {et~al.}(2014)\citenamefont {Unger},
  \citenamefont {Hoke}, \citenamefont {Bailie}, \citenamefont {Nguyen},
  \citenamefont {Bowring}, \citenamefont {Heumuller}, \citenamefont
  {Christoforo},\ and\ \citenamefont {McGehee}}]{Unger14}%
  \BibitemOpen
  \bibfield  {author} {\bibinfo {author} {\bibfnamefont {E.~L.}\ \bibnamefont
  {Unger}}, \bibinfo {author} {\bibfnamefont {E.~T.}\ \bibnamefont {Hoke}},
  \bibinfo {author} {\bibfnamefont {C.~D.}\ \bibnamefont {Bailie}}, \bibinfo
  {author} {\bibfnamefont {W.~H.}\ \bibnamefont {Nguyen}}, \bibinfo {author}
  {\bibfnamefont {A.~R.}\ \bibnamefont {Bowring}}, \bibinfo {author}
  {\bibfnamefont {T.}~\bibnamefont {Heumuller}}, \bibinfo {author}
  {\bibfnamefont {M.~G.}\ \bibnamefont {Christoforo}}, \ and\ \bibinfo {author}
  {\bibfnamefont {M.~D.}\ \bibnamefont {McGehee}},\ }\href {\doibase
  10.1039/C4EE02465F} {\bibfield  {journal} {\bibinfo  {journal} {Energy
  Environ. Sci.}\ }\textbf {\bibinfo {volume} {7}},\ \bibinfo {pages} {3690}
  (\bibinfo {year} {2014})}\BibitemShut {NoStop}%
\bibitem [{\citenamefont {Zhang}\ \emph {et~al.}(2015)\citenamefont {Zhang},
  \citenamefont {Liu}, \citenamefont {Eperon}, \citenamefont {Leijtens},
  \citenamefont {McMeekin}, \citenamefont {Saliba}, \citenamefont {Zhang},
  \citenamefont {de~Bastiani}, \citenamefont {Petrozza}, \citenamefont {Herz},
  \citenamefont {Johnston}, \citenamefont {Lin},\ and\ \citenamefont
  {Snaith}}]{Zhang_15}%
  \BibitemOpen
  \bibfield  {author} {\bibinfo {author} {\bibfnamefont {Y.}~\bibnamefont
  {Zhang}}, \bibinfo {author} {\bibfnamefont {M.}~\bibnamefont {Liu}}, \bibinfo
  {author} {\bibfnamefont {G.~E.}\ \bibnamefont {Eperon}}, \bibinfo {author}
  {\bibfnamefont {T.~C.}\ \bibnamefont {Leijtens}}, \bibinfo {author}
  {\bibfnamefont {D.}~\bibnamefont {McMeekin}}, \bibinfo {author}
  {\bibfnamefont {M.}~\bibnamefont {Saliba}}, \bibinfo {author} {\bibfnamefont
  {W.}~\bibnamefont {Zhang}}, \bibinfo {author} {\bibfnamefont
  {M.}~\bibnamefont {de~Bastiani}}, \bibinfo {author} {\bibfnamefont
  {A.}~\bibnamefont {Petrozza}}, \bibinfo {author} {\bibfnamefont {L.~M.}\
  \bibnamefont {Herz}}, \bibinfo {author} {\bibfnamefont {M.~B.}\ \bibnamefont
  {Johnston}}, \bibinfo {author} {\bibfnamefont {H.}~\bibnamefont {Lin}}, \
  and\ \bibinfo {author} {\bibfnamefont {H.~J.}\ \bibnamefont {Snaith}},\
  }\href {\doibase 10.1039/C4MH00238E} {\bibfield  {journal} {\bibinfo
  {journal} {Mater. Horiz.}\ }\textbf {\bibinfo {volume} {2}},\ \bibinfo
  {pages} {315} (\bibinfo {year} {2015})}\BibitemShut {NoStop}%
\bibitem [{\citenamefont {Chen}\ \emph {et~al.}(2016)\citenamefont {Chen},
  \citenamefont {Yang}, \citenamefont {Priya},\ and\ \citenamefont
  {Zhu}}]{Chen16}%
  \BibitemOpen
  \bibfield  {author} {\bibinfo {author} {\bibfnamefont {B.}~\bibnamefont
  {Chen}}, \bibinfo {author} {\bibfnamefont {M.}~\bibnamefont {Yang}}, \bibinfo
  {author} {\bibfnamefont {S.}~\bibnamefont {Priya}}, \ and\ \bibinfo {author}
  {\bibfnamefont {K.}~\bibnamefont {Zhu}},\ }\href@noop {} {\bibfield
  {journal} {\bibinfo  {journal} {J. Phys. Chem. Lett.}\ }\textbf {\bibinfo
  {volume} {7}},\ \bibinfo {pages} {905} (\bibinfo {year} {2016})}\BibitemShut
  {NoStop}%
\bibitem [{\citenamefont {Jena}\ \emph {et~al.}(2016)\citenamefont {Jena},
  \citenamefont {Kulkarni}, \citenamefont {Ikegami},\ and\ \citenamefont
  {Miyasaka}}]{Jena16}%
  \BibitemOpen
  \bibfield  {author} {\bibinfo {author} {\bibfnamefont {A.~K.}\ \bibnamefont
  {Jena}}, \bibinfo {author} {\bibfnamefont {A.}~\bibnamefont {Kulkarni}},
  \bibinfo {author} {\bibfnamefont {M.}~\bibnamefont {Ikegami}}, \ and\
  \bibinfo {author} {\bibfnamefont {T.}~\bibnamefont {Miyasaka}},\ }\href
  {\doibase http://dx.doi.org/10.1016/j.jpowsour.2016.01.094} {\bibfield
  {journal} {\bibinfo  {journal} {J. Power Sources}\ }\textbf {\bibinfo
  {volume} {309}},\ \bibinfo {pages} {1 } (\bibinfo {year} {2016})}\BibitemShut
  {NoStop}%
\bibitem [{\citenamefont {Cojocaru}\ \emph {et~al.}(2015)\citenamefont
  {Cojocaru}, \citenamefont {Uchida}, \citenamefont {Jayaweera}, \citenamefont
  {Kaneko}, \citenamefont {Nakazaki}, \citenamefont {Kubo},\ and\ \citenamefont
  {Segawa}}]{Cojocaru15}%
  \BibitemOpen
  \bibfield  {author} {\bibinfo {author} {\bibfnamefont {L.}~\bibnamefont
  {Cojocaru}}, \bibinfo {author} {\bibfnamefont {S.}~\bibnamefont {Uchida}},
  \bibinfo {author} {\bibfnamefont {P.~V.~V.}\ \bibnamefont {Jayaweera}},
  \bibinfo {author} {\bibfnamefont {S.}~\bibnamefont {Kaneko}}, \bibinfo
  {author} {\bibfnamefont {J.}~\bibnamefont {Nakazaki}}, \bibinfo {author}
  {\bibfnamefont {T.}~\bibnamefont {Kubo}}, \ and\ \bibinfo {author}
  {\bibfnamefont {H.}~\bibnamefont {Segawa}},\ }\href {\doibase
  10.1246/cl.150933} {\bibfield  {journal} {\bibinfo  {journal} {Chem. Lett.}\
  }\textbf {\bibinfo {volume} {44}},\ \bibinfo {pages} {1750} (\bibinfo {year}
  {2015})}\BibitemShut {NoStop}%
\bibitem [{\citenamefont {Ono}\ \emph {et~al.}(2015)\citenamefont {Ono},
  \citenamefont {Raga}, \citenamefont {Wang}, \citenamefont {Kato},\ and\
  \citenamefont {Qi}}]{Ono15}%
  \BibitemOpen
  \bibfield  {author} {\bibinfo {author} {\bibfnamefont {L.~K.}\ \bibnamefont
  {Ono}}, \bibinfo {author} {\bibfnamefont {S.~R.}\ \bibnamefont {Raga}},
  \bibinfo {author} {\bibfnamefont {S.}~\bibnamefont {Wang}}, \bibinfo {author}
  {\bibfnamefont {Y.}~\bibnamefont {Kato}}, \ and\ \bibinfo {author}
  {\bibfnamefont {Y.}~\bibnamefont {Qi}},\ }\href {\doibase 10.1039/C4TA04969A}
  {\bibfield  {journal} {\bibinfo  {journal} {J. Mater. Chem. A}\ }\textbf
  {\bibinfo {volume} {3}},\ \bibinfo {pages} {9074} (\bibinfo {year}
  {2015})}\BibitemShut {NoStop}%
\bibitem [{\citenamefont {Nagaoka}\ \emph {et~al.}(2015)\citenamefont
  {Nagaoka}, \citenamefont {Ma}, \citenamefont {deQuilettes}, \citenamefont
  {Vorpahl}, \citenamefont {Glaz}, \citenamefont {Colbert}, \citenamefont
  {Ziffer},\ and\ \citenamefont {Ginger}}]{Nagaoka15}%
  \BibitemOpen
  \bibfield  {author} {\bibinfo {author} {\bibfnamefont {H.}~\bibnamefont
  {Nagaoka}}, \bibinfo {author} {\bibfnamefont {F.}~\bibnamefont {Ma}},
  \bibinfo {author} {\bibfnamefont {D.~W.}\ \bibnamefont {deQuilettes}},
  \bibinfo {author} {\bibfnamefont {S.~M.}\ \bibnamefont {Vorpahl}}, \bibinfo
  {author} {\bibfnamefont {M.~S.}\ \bibnamefont {Glaz}}, \bibinfo {author}
  {\bibfnamefont {A.~E.}\ \bibnamefont {Colbert}}, \bibinfo {author}
  {\bibfnamefont {M.~E.}\ \bibnamefont {Ziffer}}, \ and\ \bibinfo {author}
  {\bibfnamefont {D.~S.}\ \bibnamefont {Ginger}},\ }\href {\doibase
  10.1021/jz502694g} {\bibfield  {journal} {\bibinfo  {journal} {J. Phys. Chem.
  Lett.}\ }\textbf {\bibinfo {volume} {6}},\ \bibinfo {pages} {669} (\bibinfo
  {year} {2015})}\BibitemShut {NoStop}%
\bibitem [{\citenamefont {Tao}\ \emph {et~al.}(2015)\citenamefont {Tao},
  \citenamefont {Neutzner}, \citenamefont {Colella}, \citenamefont {Marras},
  \citenamefont {Srimath~Kandada}, \citenamefont {Gandini}, \citenamefont
  {Bastiani}, \citenamefont {Pace}, \citenamefont {Manna}, \citenamefont
  {Caironi}, \citenamefont {Bertarelli},\ and\ \citenamefont
  {Petrozza}}]{Tao15}%
  \BibitemOpen
  \bibfield  {author} {\bibinfo {author} {\bibfnamefont {C.}~\bibnamefont
  {Tao}}, \bibinfo {author} {\bibfnamefont {S.}~\bibnamefont {Neutzner}},
  \bibinfo {author} {\bibfnamefont {L.}~\bibnamefont {Colella}}, \bibinfo
  {author} {\bibfnamefont {S.}~\bibnamefont {Marras}}, \bibinfo {author}
  {\bibfnamefont {A.~R.}\ \bibnamefont {Srimath~Kandada}}, \bibinfo {author}
  {\bibfnamefont {M.}~\bibnamefont {Gandini}}, \bibinfo {author} {\bibfnamefont
  {M.~D.}\ \bibnamefont {Bastiani}}, \bibinfo {author} {\bibfnamefont
  {G.}~\bibnamefont {Pace}}, \bibinfo {author} {\bibfnamefont {L.}~\bibnamefont
  {Manna}}, \bibinfo {author} {\bibfnamefont {M.}~\bibnamefont {Caironi}},
  \bibinfo {author} {\bibfnamefont {C.}~\bibnamefont {Bertarelli}}, \ and\
  \bibinfo {author} {\bibfnamefont {A.}~\bibnamefont {Petrozza}},\ }\href
  {\doibase 10.1039/C5EE01720C} {\bibfield  {journal} {\bibinfo  {journal}
  {Energy Environ. Sci.}\ }\textbf {\bibinfo {volume} {8}},\ \bibinfo {pages}
  {2365} (\bibinfo {year} {2015})}\BibitemShut {NoStop}%
\bibitem [{\citenamefont {Zhou}\ \emph {et~al.}(2015)\citenamefont {Zhou},
  \citenamefont {Huang}, \citenamefont {Cheng},\ and\ \citenamefont
  {Gray-Weale}}]{Zhou_15}%
  \BibitemOpen
  \bibfield  {author} {\bibinfo {author} {\bibfnamefont {Y.}~\bibnamefont
  {Zhou}}, \bibinfo {author} {\bibfnamefont {F.}~\bibnamefont {Huang}},
  \bibinfo {author} {\bibfnamefont {Y.-B.}\ \bibnamefont {Cheng}}, \ and\
  \bibinfo {author} {\bibfnamefont {A.}~\bibnamefont {Gray-Weale}},\ }\href
  {\doibase 10.1039/C5CP03352G} {\bibfield  {journal} {\bibinfo  {journal}
  {Phys. Chem. Chem. Phys.}\ }\textbf {\bibinfo {volume} {17}},\ \bibinfo
  {pages} {22604} (\bibinfo {year} {2015})}\BibitemShut {NoStop}%
\bibitem [{\citenamefont {Sze}\ and\ \citenamefont {Ng}(2006)}]{sze06}%
  \BibitemOpen
  \bibfield  {author} {\bibinfo {author} {\bibfnamefont {S.}~\bibnamefont
  {Sze}}\ and\ \bibinfo {author} {\bibfnamefont {K.}~\bibnamefont {Ng}},\
  }\href {https://books.google.co.jp/books?id=o4unkmHBHb8C} {\emph {\bibinfo
  {title} {Physics of Semiconductor Devices}}}\ (\bibinfo  {publisher}
  {Wiley},\ \bibinfo {year} {2006})\BibitemShut {NoStop}%
\bibitem [{\citenamefont {Liu}\ \emph {et~al.}(2015)\citenamefont {Liu},
  \citenamefont {Fan}, \citenamefont {Zhang}, \citenamefont {Shen},
  \citenamefont {Yang},\ and\ \citenamefont {Mai}}]{Liu15}%
  \BibitemOpen
  \bibfield  {author} {\bibinfo {author} {\bibfnamefont {C.}~\bibnamefont
  {Liu}}, \bibinfo {author} {\bibfnamefont {J.}~\bibnamefont {Fan}}, \bibinfo
  {author} {\bibfnamefont {X.}~\bibnamefont {Zhang}}, \bibinfo {author}
  {\bibfnamefont {Y.}~\bibnamefont {Shen}}, \bibinfo {author} {\bibfnamefont
  {L.}~\bibnamefont {Yang}}, \ and\ \bibinfo {author} {\bibfnamefont
  {Y.}~\bibnamefont {Mai}},\ }\href {\doibase 10.1021/acsami.5b00375}
  {\bibfield  {journal} {\bibinfo  {journal} {ACS Appl. Mater. Interfaces}\
  }\textbf {\bibinfo {volume} {7}},\ \bibinfo {pages} {9066} (\bibinfo {year}
  {2015})}\BibitemShut {NoStop}%
\bibitem [{\citenamefont {Kim}\ and\ \citenamefont {Park}(2014)}]{Kim_14}%
  \BibitemOpen
  \bibfield  {author} {\bibinfo {author} {\bibfnamefont {H.-S.}\ \bibnamefont
  {Kim}}\ and\ \bibinfo {author} {\bibfnamefont {N.-G.}\ \bibnamefont {Park}},\
  }\href {\doibase 10.1021/jz501392m} {\bibfield  {journal} {\bibinfo
  {journal} {J. Phys. Chem. Lett.}\ }\textbf {\bibinfo {volume} {5}},\ \bibinfo
  {pages} {2927} (\bibinfo {year} {2014})}\BibitemShut {NoStop}%
\bibitem [{SI()}]{SI}%
  \BibitemOpen
  \href@noop {} {}\bibinfo {note} {See supplementary material at [URL will be
  inserted by AIP] for for an improved comparison to experiment.}\BibitemShut
  {Stop}%
\bibitem [{\citenamefont {Shao}\ \emph {et~al.}(2014)\citenamefont {Shao},
  \citenamefont {Xiao}, \citenamefont {Bi}, \citenamefont {Yuan},\ and\
  \citenamefont {Huang}}]{Shao14}%
  \BibitemOpen
  \bibfield  {author} {\bibinfo {author} {\bibfnamefont {Y.}~\bibnamefont
  {Shao}}, \bibinfo {author} {\bibfnamefont {Z.}~\bibnamefont {Xiao}}, \bibinfo
  {author} {\bibfnamefont {C.}~\bibnamefont {Bi}}, \bibinfo {author}
  {\bibfnamefont {Y.}~\bibnamefont {Yuan}}, \ and\ \bibinfo {author}
  {\bibfnamefont {J.}~\bibnamefont {Huang}},\ }\href {\doibase
  10.1038/ncomms6784} {\bibfield  {journal} {\bibinfo  {journal} {Nat Commun}\
  }\textbf {\bibinfo {volume} {5}},\ \bibinfo {pages} {10.1038/ncomms6784}
  (\bibinfo {year} {2014})}\BibitemShut {NoStop}%
\bibitem [{\citenamefont {Tripathi}\ \emph {et~al.}(2015)\citenamefont
  {Tripathi}, \citenamefont {Yanagida}, \citenamefont {Shirai}, \citenamefont
  {Masuda}, \citenamefont {Han},\ and\ \citenamefont {Miyano}}]{Tripathi15}%
  \BibitemOpen
  \bibfield  {author} {\bibinfo {author} {\bibfnamefont {N.}~\bibnamefont
  {Tripathi}}, \bibinfo {author} {\bibfnamefont {M.}~\bibnamefont {Yanagida}},
  \bibinfo {author} {\bibfnamefont {Y.}~\bibnamefont {Shirai}}, \bibinfo
  {author} {\bibfnamefont {T.}~\bibnamefont {Masuda}}, \bibinfo {author}
  {\bibfnamefont {L.}~\bibnamefont {Han}}, \ and\ \bibinfo {author}
  {\bibfnamefont {K.}~\bibnamefont {Miyano}},\ }\href {\doibase
  10.1039/C5TA01668A} {\bibfield  {journal} {\bibinfo  {journal} {J. Mater.
  Chem. A}\ }\textbf {\bibinfo {volume} {3}},\ \bibinfo {pages} {12081}
  (\bibinfo {year} {2015})}\BibitemShut {NoStop}%
\bibitem [{\citenamefont {Zarazua}\ \emph {et~al.}(2016)\citenamefont
  {Zarazua}, \citenamefont {Bisquert},\ and\ \citenamefont
  {Garcia-Belmonte}}]{Zarazua16}%
  \BibitemOpen
  \bibfield  {author} {\bibinfo {author} {\bibfnamefont {I.}~\bibnamefont
  {Zarazua}}, \bibinfo {author} {\bibfnamefont {J.}~\bibnamefont {Bisquert}}, \
  and\ \bibinfo {author} {\bibfnamefont {G.}~\bibnamefont {Garcia-Belmonte}},\
  }\href {\doibase 10.1021/acs.jpclett.5b02810} {\bibfield  {journal} {\bibinfo
   {journal} {J. Phys. Chem. Lett.}\ }\textbf {\bibinfo {volume} {7}},\
  \bibinfo {pages} {525} (\bibinfo {year} {2016})}\BibitemShut {NoStop}%
\bibitem [{\citenamefont {Wu}\ \emph {et~al.}(2015)\citenamefont {Wu},
  \citenamefont {Fu}, \citenamefont {Yantara}, \citenamefont {Xing},
  \citenamefont {Sun}, \citenamefont {Sum},\ and\ \citenamefont
  {Mathews}}]{Bo15}%
  \BibitemOpen
  \bibfield  {author} {\bibinfo {author} {\bibfnamefont {B.}~\bibnamefont
  {Wu}}, \bibinfo {author} {\bibfnamefont {K.}~\bibnamefont {Fu}}, \bibinfo
  {author} {\bibfnamefont {N.}~\bibnamefont {Yantara}}, \bibinfo {author}
  {\bibfnamefont {G.}~\bibnamefont {Xing}}, \bibinfo {author} {\bibfnamefont
  {S.}~\bibnamefont {Sun}}, \bibinfo {author} {\bibfnamefont {T.~C.}\
  \bibnamefont {Sum}}, \ and\ \bibinfo {author} {\bibfnamefont
  {N.}~\bibnamefont {Mathews}},\ }\href {\doibase 10.1002/aenm.201500829}
  {\bibfield  {journal} {\bibinfo  {journal} {Adv. Energy Mater.}\ }\textbf
  {\bibinfo {volume} {5}},\ \bibinfo {pages} {1500829} (\bibinfo {year}
  {2015})}\BibitemShut {NoStop}%
\end{thebibliography}
%


\end{document}